\documentclass[11pt,twoside]{article}

\leftline{\copyright~ 1998 International Press}
\leftline{Adv. Theor. Math. Phys. {\bf 2} (1998) {\ 963 -- 985} }

\usepackage{verbatim}

\newtheorem{Def}{Def.}[section]
\newtheorem{Thm}[Def]{Theorem}

\newcommand{\Proof}{{\em{Proof: }}}
\newcommand{\QED}{\ \hfill $\FBox$ \\[1em]}
\newcommand{\Equ}[1]{\begin{equation} \label{eq:#1}}
\newcommand{\EndEqu}{\end{equation}}

\newcommand{\Pdd}{\mbox{$\partial$ \hspace{-1.2 em} $/$}}

\newcommand{\bra}{\mbox{$< \!\!$ \nolinebreak}}
\newcommand{\ket}{\mbox{\nolinebreak $>$}}

\newcommand{\spc}{\;\;\;\;\;\;\;\;\;\;}

\newcommand{\Aslsh}{\mbox{ $\!\!A$ \hspace{-1.2 em} $/$}}
\newcommand{\slsh}{\mbox{ \hspace{-1.1 em} $/$}}

\newcommand{\1}{\mbox{\rm 1 \hspace{-1.05 em} 1}}

\newcommand{\sR}{\mbox{\rm \scriptsize I \hspace{-.8 em} R}}
\newcommand{\N}{\mbox{\rm I \hspace{-.8 em} N}}

\newcommand{\FBox}{\rule{2mm}{2.25mm}}

\begin{document}

\begin{center}
\vspace{0.4in}

{\huge \bf Definition of the Dirac Sea in \\[5mm] the Presence of External Fields} 
\renewcommand{\thefootnote}{}
\footnotetext{e-print archive: {\texttt http://xxx.lanl.gov/abs/hep-th/9705006}}
\renewcommand{\thefootnote}{\arabic{footnote}}

\vspace{0.4in}
 {\bf Felix Finster} \footnote{Supported by the Deutsche Forschungsgemeinschaft, Bonn.}

\vspace{0.2in}

Mathematics Department \\
Harvard University \\




\begin{abstract}
	It is shown that the Dirac sea can be uniquely defined for the Dirac 
	equation with general interaction, if we impose a causality 
	condition on the Dirac sea. We derive an explicit formula for the 
	Dirac sea in terms of a power series in the bosonic potentials.
	
	The construction is extended to systems of Dirac seas. If the system 
	contains chiral fermions, the causality condition yields a 
	restriction for the bosonic potentials.
\end{abstract}
\end{center}

\setcounter{page}{964} 

\pagenumbering{arabic}

\section{Introduction}
The Dirac equation has solutions of negative energy, which have no
meaningful
physical interpretation. This popular problem of relativistic quantum
mechanics
was originally solved by Dirac's
concept that all negative-energy states are occupied in the vacuum
forming
the so-called Dirac sea. Fermions and anti-fermions are then described
by
positive-energy states and ``holes'' in the Dirac sea, respectively.
Although this vivid picture of a sea of interacting particles is
nowadays
often considered not to be taken too literally, the construction of the 
Dirac sea also plays a crucial role in quantum field theory.
There it corresponds to the formal exchanging of creation and
annihilation 
operators for the negative-energy states of the free field theory.
\newpage
Usually, the Dirac sea is only constructed in the vacuum.
This is often considered to be sufficient, because the interacting
system can
be described by a perturbation of the vacuum.

Unfortunately, the situation is more difficult: In relativistic quantum
mechanics
with interaction, the fermionic wave functions are solutions of the
Dirac equation
\begin{equation}
(i \Pdd + {\cal{B}} - m) \:\tilde{\Psi} \;=\; 0,
\label{1}
\end{equation}

\pagestyle{myheadings}
\markboth{\it DEFINITION OF THE DIRAC SEA ...}{\it F. FINSTER}
\setcounter{page}{964}

where the operator ${\cal{B}}$ is composed of the bosonic potentials
(for example, we can describe the electromagnetic interaction by
choosing
${\cal{B}}=e \Aslsh$ with the electromagnetic potential $A$).

In contrast to the free Dirac equation $(i \Pdd - m) \:\Psi =0$, it is
not obvious
how to characterize the negative-energy solutions of (\ref{1}).
Qualitatively, the
problem is that the perturbation ${\cal{B}}$ leads to a mixing of the
free
solutions and destroys the natural splitting into solutions of
positive and negative energy. As a consequence, it is not clear how the
Dirac sea
of the system (\ref{1}) can be constructed. We point out that this
problem is
not solved by a simple perturbation expansion in ${\cal{B}}$; it is then
hidden
in the non-uniqueness of this expansion (see section \ref{sec2} for
details).
In quantum field theory, the problem
of defining the Dirac sea is even more complicated, because the 
virtual pair creation/annihilation must be taken into account.
We will not deal these problems here and restrict to the limit of
``classical'' potentials and wave functions. Nevertheless, our 
considerations are also relevant for quantum field theory, because
it is in many situations (e.g.\ for a quantum system in a classical
background field) preferable to use the Dirac equation (\ref{1}) as
the starting point for the fermionic field quantization.
In this sense, the construction of the Dirac sea of (\ref{1}) is
preliminary
for the description of interacting quantum fields.

We conclude that the definition of the Dirac sea is basic for a
reasonable
physical interpretation of the Dirac equation (\ref{1}).
In the present paper, we will discuss the difficulty in constructing 
the Dirac sea and finally solve the problem in terms of a formal 
perturbation expansion in ${\cal{B}}$.
Before starting the analysis, we describe the problem in more
mathematical
terms: Every solution of the free Dirac equation $(i \Pdd - m) 
\:\Psi=0$ is a linear combination of plane wave solutions of the form
\[ \Psi(t, \vec{x}) \;=\; e^{-i(\omega t - \vec{k} \vec{x})} 
\:\chi_{\omega, \vec{k}} \;\;\;,\spc \omega \;=\; \pm 
\sqrt{\vec{k}^2 + m^2} \]
with a $4$-spinor $\chi_{\omega, \vec{k}}$ which is independent of $t$ 
and $\vec{x}$. The sign of $\omega$ gives a natural splitting of the 
solutions into solutions of positive and negative frequency. 
Identifying frequency and energy via Planck's formula, these solutions
are commonly called the positive and 
negative energy solutions of the free Dirac equation. Since the simple 
identification of frequency and energy might lead to confusion 
(sometimes the ``energy'' of a negative-frequency state denotes the 
positive energy of the corresponding anti-particle state), we prefer 
the notion of positive and negative ``frequency'' in the following.
We denote the negative-frequency solutions by $\Psi_{\vec{k} a}$,
where $\vec{k}$ is the momentum and $a=1,2$ are the two spin states
(for an explicit formula for $\Psi_{\vec{k}a}$ see e.g.\ \cite{BD}).
If the states $\Psi_{\vec{k}a}$ were normalized with respect to the 
usual scalar product

\begin{equation}
	(\Psi \:|\: \Phi) \;=\; \int_{\sR^3} (\overline{\Psi} \:\gamma^0\: 
	\Phi)(t,\vec{x}) \; d\vec{x} \;\;\;,\spc \overline{\Psi}=\Psi^* 
	\gamma^0 \spc ,
	\label{s0}
\end{equation}
we could form the projector $P_{\bra \Psi_{\vec{k}a} \ket}$ on the
one-dimensional subspace 
$\bra \Psi_{\vec{k}a} \ket$ by
\[ \left( P_{\bra \Psi_{\vec{k}a} \ket} \:\Psi \right)(t,\vec{x}) 
\;=\; \int_{\sR^3} \left( \Psi_{\vec{k}a}(t,\vec{x}) \: 
\overline{\Psi_{\vec{k}a}(t,\vec{y})} \right) \: \gamma^0 \:
\Psi(t,\vec{y}) 
\; d\vec{y} \spc . \]
In this sense, the product
$\Psi_{\vec{k} a}(x) \: \overline{\Psi_{\vec{k} a}(y)}$ would be the
kernel of the projector on $\bra \Psi_{\vec{k}a} \ket$, and the sum over
all 
negative-frequency states would yield the projector on the whole Dirac
sea.
Unfortunately, the wave functions $\Psi_{\vec{k}a}$ are not 
normalizable. We could arrange normalizable states by considering 
the system in finite three-volume, but we do not want to do this here.
It is more appropriate for our purpose to formally build up a projector
on 
all negative-frequency states by integrating over the momentum parameter
\begin{equation}
P(x,y) \;=\; \sum_{a=1,2} \int_{\sR^3} \: \Psi_{\vec{k} a}(x) \:
	\overline{\Psi_{\vec{k} a}(y)} \; d\vec{k} \label{2a} \spc ,
\end{equation}
which can be rewritten as the integral over the lower mass shell
\[ \;=\; \int_{\sR^3} \frac{d^4k}{(2 \pi)^4} \: (k \slsh + m) \:
	\delta(k^2-m^2) \: \Theta(-k^0) \: e^{-ik(x-y)} \]
($\Theta$ denotes the Heavyside function $\Theta(x)=1$ for $x \geq 0$ 
and $\Theta(x)=0$ otherwise).
$P(x,y)$ is a well-defined tempered distribution which solves the
free Dirac equation $(i \Pdd_{x} - m) \:P(x,y) = 0$. We can use it to
characterize the Dirac sea in the vacuum.
Our aim is to introduce a corresponding distribution $\tilde{P}$ for the
Dirac equation with interaction (\ref{1}).
The construction of $\tilde{P}$ must be unique in a sense which we 
will discuss and specify later.
We will assume the perturbation ${\cal{B}}$ to be a differential
operator on
the wave functions.
Furthermore, it shall be Hermitian with respect to the (indefinite)
scalar product
\begin{equation}
\bra \Psi \:|\: \Phi \ket \;=\; \int \overline{\Psi(x)} \: \Phi(x) \;
d^4x
\spc .
\label{3z}
\end{equation}
For an electromagnetic potential ${\cal{B}}=e \Aslsh$, these assumptions 
are satisfied because $\Aslsh=\gamma^{0} \Aslsh^{\dagger} \gamma^{0}$.
In addition, ${\cal{B}}$ can be composed of the scalar, pseudoscalar, 
pseudovector and bilinear potentials as e.g.\ discussed in \cite{T}.
According to \cite{F2}, ${\cal{B}}$ also allows for the description
of the gravitational field.

\section{Non-Uniqueness of the Simple Perturbation Expansion}
\label{sec2}
\setcounter{equation}{0}
Our first idea for the construction of $\tilde{P}$ is to 
calculate solutions $\tilde{\Psi}_{\vec{k} a}$ of (\ref{1}) with a 
perturbation expansion in ${\cal{B}}$ and to define $\tilde{P}$ in
analogy to
(\ref{2a}) by
\begin{equation}
	\tilde{P}(x,y) \;=\; \sum_{a=1,2} \int_{\sR^3} \tilde{\Psi}_{\vec{k} 
	a}(x) \: \overline{\tilde{\Psi}_{\vec{k} a}(y)} \: d\vec{k} \spc .
	\label{4y}
\end{equation}
We start with a discussion of this method in a perturbation 
calculation to first order. This is quite elementary and will 
nevertheless explain the basic difficulty.
For the perturbation calculation, we need a Green's function 
$s(x,y)$ of the free Dirac operator, which is
characterized by the distributional equation
\begin{equation}
(i \Pdd_x - m) \: s(x,y) \;=\; \delta^4(x-y) \spc . \label{4}
\end{equation}
To first order, the perturbed eigenstates $\tilde{\Psi}_{\vec{k} a}$ 
are then given by
\begin{equation}
\tilde{\Psi}_{\vec{k} a}(x) \;=\; \Psi_{\vec{k} a}(x) \:-\: \int d^4y \;
	s(x,y) \: {\cal{B}}_y \: \Psi_{\vec{k} a}(y) \;+\;
{\cal{O}}({\cal{B}}^2)
	\spc ,
\label{5}
\end{equation}
as can be verified by substituting into (\ref{1}).
We insert this formula into (\ref{4y}) and obtain
\begin{equation}
	\tilde{P}(x,y) \;=\; P(x,y) \:-\: \int d^{4}z \; \left[ s(x,z) \:
	{\cal{B}}_{z} \: P(z,y) \:+\: P(x,z) \:{\cal{B}}_{z}\: s^{*}(z,y) 
	\right] \:+\: {\cal{O}}({\cal{B}}^{2}) ,
	\label{7}
\end{equation}
where we used that ${\cal{B}}$ is Hermitian with respect to the scalar 
product (\ref{3z}), and where $s^{*}(z,y)$ is given by $s^{*}(z,y) =
\gamma^{0}\:s(y,z)^{\dagger} \:\gamma^{0}$.
It is convenient to view the distributions $s(x,y), P(x,y)$ as integral 
kernels of corresponding operators $s, P$. Then we can write (\ref{7})
with
operator products
\begin{equation}
	\tilde{P} \;=\; P \:-\: s \:{\cal{B}}\: P \:-\: P 
	\:{\cal{B}}\: s^{*} \:+\: {\cal{O}}({\cal{B}}^{2}) \spc ,
	\label{8}
\end{equation}
where the superscript `$^{*}$' denotes the adjoint with respect to 
the scalar product (\ref{3z}).

Equation (\ref{8}) gives a possible definition for $\tilde{P}$.
As apparent problem, the construction depends on the choice of the 
Green's function. For example, we could have chosen for $s$ either the
advanced or 
the retarded Green's function $s^{\vee}_{m}$,$s^{\wedge}_{m}$, which are
in 
momentum space as usual given by
\begin{equation}
	s^{\vee}_{m}(k) \;=\; \lim_{0<\varepsilon \rightarrow 0}
	    \frac{k \slsh + m}{k^{2}-m^{2}-i \varepsilon k^{0}} 
	    \;\;\;,\;\;\;\;\;
	   s^{\wedge}_{m}(k) \;=\; \lim_{0<\varepsilon \rightarrow 0}
	    \frac{k \slsh + m}{k^{2}-m^{2}+i \varepsilon k^{0}}  
	\label{8b}
\end{equation}
More systematically, the arbitrariness of our construction is
described as follows:
According to (\ref{4}), the difference between two Green's functions 
is a solution of the free Dirac equation. We can thus represent $s$ in 
the form
\[ s(x,y) \;=\; s^{\vee}_{m}(x,y) \:+\: a(x,y) \spc , \]
where $a(x,y)$ is in the $x$-variable a linear combination of the
plane-wave solutions, i.e.
\[ a(x,y) \;=\; \sum_{a=1}^{4} \int_{\sR^3} \Psi_{\vec{k} a}(x) \:
c_{\vec{k} a}(y) \; d\vec{k} \]
with (generally complex) functions $c_{\vec{k} a}(y)$, where
$\Psi_{\vec{k}a}, a=3,4$
denote the plane-wave solutions of positive frequency. We substitute 
into (\ref{8}) and obtain
\begin{equation}
	\tilde{P} \;=\; P \:-\: s^{\vee}_{m}\:{\cal{B}}\:P \:-\: 
	P\:{\cal{B}}\:s^{\wedge}_{m} \:-\: \left( a \:{\cal{B}}\: P \:+\: 
	P \:{\cal{B}} \: a^{*} \right) \;+\; {\cal{O}}({\cal{B}}^{2}) \spc .
	\label{9}
\end{equation}
The expression in the brackets maps solutions of the free Dirac 
equation into each other and vanishes otherwise. We can thus write it 
in the form
\begin{equation}
(a \:{\cal{B}}\: P \:+\: P \:{\cal{B}} \: a^{*})(x,y)
\;=\; \sum_{a,b=1}^{4} \int_{\sR^3} d\vec{k}_{1} \int_{\sR^3}
d\vec{k}_{2} \; 
\Psi_{\vec{k}_{1} a}(x) \: g_{ab}(\vec{k}_{1}, \vec{k}_{2}) \:
\overline{ \Psi_{\vec{k}_{2} b}(y) }
	\label{10}
\end{equation}
with suitable functions $g_{ab}(\vec{k}_{1}, \vec{k}_{2})$.
This representation of $\tilde{P}$ can also be understood directly:
The contribution (\ref{10}) describes a mixing of the solutions 
$\Psi_{\vec{k}a}$ of the free Dirac equation. To the considered first 
order in ${\cal{B}}$, it vanishes in the Dirac 
equation $(i \Pdd + {\cal{B}} - m) \:\tilde{P}=0$. Thus we cannot fix 
this contribution with the Dirac equation, it remains undetermined 
in our method. According to (\ref{9}), this is the only 
arbitrariness of the construction; the other contributions to 
$\tilde{P}$ are unique.

In higher order perturbation theory, the non-uniqueness can 
be understood similarly, although the situation is more complicated:
For a given Green's function $s$, we can construct a 
solution $\tilde{\Psi}_{\vec{k} a}$ of the Dirac equation (\ref{1})
by the formal perturbation series
\begin{equation}
	\tilde{\Psi}_{\vec{k}a} \;=\; \sum_{n=0}^{\infty} (-s \: 
	{\cal{B}})^{n} \: \Psi_{\vec{k}a} \spc ,
	\label{11}
\end{equation}
as is verified by substituting into (\ref{1}).
Actually, this is a very special ansatz. For example, we can use
different Green's functions in every order of the perturbation 
calculation, which leads to the more general formula
\begin{equation}
\tilde{\Psi}_{\vec{k}a} \;=\; \Psi_{\vec{k}a} \:+\:
	\sum_{n=1}^{\infty} (-1)^{n} \; s^{(n)} \:{\cal{B}}\: \cdots \:
	s^{(2)} \:{\cal{B}}\: s^{(1)} \:{\cal{B}} \:\Psi_{\vec{k}a}
	\label{12}
\end{equation}
with a whole series of arbitrary Green's functions 
$s^{(1)}$, $s^{(2)}$, etc.. Once we have a formula for
$\tilde{\Psi}_{\vec{k}a}$, the non-uniqueness of $\tilde{P}$ can again
be
discussed by substituting into (\ref{4y}).
In generalization of (\ref{10}), the arbitrariness of the construction 
is described by a contribution to $\tilde{P}(x,y)$ of the form
\[ \sum_{a,b=1}^{4} \int_{\sR^3} d\vec{k}_{1} \int_{\sR^3} d\vec{k}_{2}
\;
\tilde{\Psi}_{\vec{k}_{1}a}(x) \:g_{ab}(\vec{k}_{1}, \vec{k}_{2}) \:
\overline{ \tilde{\Psi}_{\vec{k}_{2}b}(y) } \spc , \]
which mixes perturbed eigenstates $\tilde{\Psi}_{\vec{k}a}$ and 
vanishes in the Dirac equation $(i \Pdd + {\cal{B}} - m) \:\tilde{P}=0$.
The dependence of $g_{ab}(\vec{k}_{1}, \vec{k}_{2})$ on ${\cal{B}}$ 
and on the Green's functions $s^{(n)}$ is rather involved, however, and
we 
need not go into the details here.

To summarize, a simple perturbation expansion in ${\cal{B}}$ is not
unique 
and therefore does not allow a meaningful definition of $\tilde{P}$.
In the ansatz (\ref{12}), for example, we should
find a way to specify the Green's functions $s^{(n)}$. This 
cannot be done with the Dirac equation (\ref{1}), and we must therefore
look for additional input to completely determine $\tilde{P}$.
Our basic idea is to apply some causality principle. For example, it 
might seem a reasonable condition to impose that $\tilde{P}(x,y)$ only 
depends on ${\cal{B}}$ in the ``diamond'' $(L^{\vee}_{x} \cap 
L^{\wedge}_{y}) \:\cup\: (L^{\wedge}_{x} \cap L^{\vee}_{y})$, where
\begin{eqnarray}
 L^{\vee}_{x} \;=\; \left\{ y \:|\: (y-x)^{2} \geq 0 ,\;
     y^{0}-x^{0} \geq 0 \right\} \\
 L^{\wedge}_{x} \;=\; \left\{ y \:|\: (y-x)^{2} \geq 0 ,\;
     y^{0}-x^{0} \leq 0 \right\} 
\end{eqnarray}
denote the future and past light cones around $x$, respectively.
If we want to study conditions of this type, it is no longer useful to 
look at the perturbation expansion for the individual states 
$\Psi_{\vec{k}a}(x)$ (because these states only depend on one 
argument $x$). We must take into account for the perturbation expansion
that $P$ is composed of many states in a specific way.

\section{The Causal Perturbation Expansion}
\setcounter{equation}{0}
In preparation, we first describe how the perturbation expansion for 
the advanced and retarded Green's functions can be performed uniquely:
The support of the distribution $s^{\vee}_{m}(x,y)$ is 
in the future light cone $y \in L^{\vee}_{x}$ (this can be checked by
calculating the Fourier transform of (\ref{8b}) with contour integrals).
As a consequence, the perturbation operator ${\cal{B}}(z)$ only enters
into
the operator product
\begin{equation}
	(s^\vee_m \:{\cal{B}}\: s^\vee_m)(x,y) \;=\;
		\int d^4z \; s^\vee_m(x,z) \: {\cal{B}}(z) \: s^\vee_m(z,y)
	\label{2t1}
\end{equation}
for $z \in L^{\vee}_{x} \cap L^{\wedge}_{y}$.
In this sense, the expression (\ref{2t1}) is
{\em{causal}}. Especially, the support of (\ref{2t1}) is again in the 
future light cone. It follows by iteration that the higher powers
\[ s^\vee_m \:{\cal{B}}\: s^\vee_m \:{\cal{B}}\: \cdots {\cal{B}}\:
	s^\vee_m \:{\cal{B}}\: s^\vee_m \]
are also causal and have their support in the upper light cone. We 
define the perturbed advanced Green's function as the formal sum over
these
operator products,
\begin{equation}
	\tilde{s}_m^\vee \;=\; \sum_{k=0}^\infty \left(- s_m^\vee \: {\cal{B}}
		\right)^k \: s_m^\vee \spc .
	\label{2t2}
\end{equation}
Accordingly, the perturbed retarded Green's function is defined by
\begin{equation}
	\tilde{s}_m^\wedge \;=\; \sum_{k=0}^\infty \left(- s_m^\wedge \:
	{\cal{B}} \right)^k \: s_m^\wedge \spc .
	\label{2t3}
\end{equation}
These operators satisfy the defining equations for the perturbed Green's
functions
\begin{equation}
	(i \Pdd - m + {\cal{B}}) \: \tilde{s}^\vee_m \;=\; \1 \;=\;
	(i \Pdd - m + {\cal{B}}) \: \tilde{s}^\wedge_m \spc ,
	\label{2t4}
\end{equation}
as is verified directly.

Notice that the perturbation expansion for the Green's functions 
becomes unique by the condition that the contribution to
$\tilde{s}^\vee_m,
\tilde{s}^\wedge_m$ to every order has its support in the future and
past
light cones, respectively. We want to take this construction as the
guiding
line for the perturbation expansion of $P$.
Unfortunately, the method cannot be directly applied to the Dirac sea,
because the distribution $P(x,y)$ does not vanish for space-like 
$y-x$, and we thus have no notion of causality.
As way out, we decompose the free Dirac sea in the form
\begin{equation}
	P(x,y) \;=\; \frac{1}{2} \left( p_{m}(x,y) \:-\: k_{m}(x,y) \right)
	\label{12b}
\end{equation}
with the tempered distributions
\begin{eqnarray}
p_{m}(x,y) &=& \int \frac{d^{4}k}{(2 \pi)^{4}} \:(k \slsh + m) \:
  \delta(k^{2}-m^{2}) \: e^{-ik(x-y)}
  \label{20} \\
k_{m}(x,y) &=& \int \frac{d^{4}k}{(2 \pi)^{4}} \:(k \slsh + m) \:
  \delta(k^{2}-m^{2}) \:\epsilon(k^{0}) \: e^{-ik(x-y)}
  \label{21}
\end{eqnarray}
($\epsilon$ denotes the step function $\epsilon(x)=1$ for $x \geq 0$ 
and $\epsilon(x)=-1$ otherwise).
We also consider $p_{m}(x,y)$ and $k_{m}(x,y)$ as integral kernels of 
corresponding operators $p_{m}$, $k_{m}$.
The operator $p_{m}$ is built up as a formal sum over the projectors on 
all solutions of the Dirac equation and can be viewed as a spectral
projector
of the free Dirac operator. The definition of $k_{m}$
differs from $p_{m}$ by a relative minus sign for the states on 
the upper and lower mass shell.
As a consequence of this relative minus sign, the Fourier integral
(\ref{21})
vanishes if $y-x$ is space-like (this can be seen from Lorentzian
invariance
and a symmetry argument for $k=(0,\vec{k})$). Thus $k_{m}(x,y)$ is
causal in 
the sense that it has the support in the light cone $y \in 
L^{\vee}_{x} \cup L^{\wedge}_{x}$.
This makes it possible to uniquely express 
its perturbation expansion in terms of the perturbed Green's functions:
We substitute the distributional equation
\[ \lim_{0<\varepsilon \rightarrow 0} \left( \frac{1}{x-i\varepsilon}
\:-\: \frac{1}{x+i \varepsilon} \right) \;=\; 2 \pi i \: \delta(x) \]
into the formula for $k_{m}$ in momentum space,
\begin{eqnarray*}
k_{m}(p) &=& (p \slsh + m) \: \delta(p^{2}-m^{2}) \: \epsilon(p^{0}) \\
&=& \frac{1}{2 \pi i} \:(p \slsh + m) \; \lim_{0<\varepsilon 
\rightarrow 0} \left[ \frac{1}{p^{2}-m^{2}-i\varepsilon} \:-\:
\frac{1}{p^{2}-m^{2}+i\varepsilon} \right] \: \epsilon(p^{0}) \\
&=& \frac{1}{2 \pi i} \:(p \slsh + m) \; \lim_{0<\varepsilon 
\rightarrow 0} \left[ \frac{1}{p^{2}-m^{2}-i\varepsilon p^{0}} \:-\:
\frac{1}{p^{2}-m^{2}+i\varepsilon p^{0}} \right] \spc ,
\end{eqnarray*}
and obtain with (\ref{8b}) a simple relation between $k_{m}$ and
$s^{\vee}_{m}$, $s^{\wedge}_{m}$,
\begin{equation}
	k_{m} \;=\; \frac{1}{2\pi i} \: \left( s^{\vee}_{m} \:-\:
s^{\wedge}_{m}
	\right) \spc .
	\label{32}
\end{equation}
We extend this relation to the case with external fields:
\begin{Def}
\label{2_def7}
We define the operator $\tilde{k}_{m}$ by
\begin{eqnarray}
\label{2tm}
\tilde{k}_m &=& \frac{1}{2 \pi i} \: \left(\tilde{s}_m^\vee -
	\tilde{s}_m^\wedge \right)
\end{eqnarray}
with the Green's functions (\ref{2t2}),(\ref{2t3}).
\end{Def}
According to (\ref{2t4}), $\tilde{k}_{m}$ really is a solution 
of the Dirac equation $(i \Pdd + {\cal{B}} - m) \:\tilde{k}_{m}=0$.

In order to explain the significance of this construction, we point out
that
the factor $\epsilon(k^{0})$ in (\ref{21}) describes the splitting of 
the solutions of the free Dirac equation into solutions of positive 
and negative frequency. With the introduction of $\tilde{k}_{m}$, we 
were able to uniquely generalize this splitting to the case with
external 
fields. This solves the basic problem in defining the Dirac sea.
It remains to perform the perturbation expansion for $\tilde{p}_m$. On a
formal level, this is very easy, because we can remove the relative
minus
sign for the positive and negative frequency states by taking the
absolute
value of $\tilde{k}_m$,
\begin{equation}
	\tilde{p}_m \;\stackrel{\mbox{\scriptsize{formally}}}{:=}\;
	\sqrt{\tilde{k}_m^2} \spc .
	\label{51}
\end{equation}
This gives a unique definition for $\tilde{p}_m$. Since 
$\tilde{k}_m$ is composed of eigenstates of the perturbed Dirac 
operator with eigenvalue $m$, it follows automatically that
$(i \Pdd + {\cal{B}} -  m) \:\tilde{p}_m = 0$.

Unfortunately, it requires some effort to convert the formal relation
(\ref{51}) into a mathematical definition. The problem is that the
square 
$\tilde{k}_m^2$ is ill-defined; furthermore we want to write 
$\tilde{p}_m$ as a power series in ${\cal{B}}$. These problems 
are solved in the following theorem. The reader who is not so 
interested in the technical details and the combinatorics of the 
expansion may skip the proof. For the statement of the theorem, we 
need some notation: We work with the Green's function
\begin{equation}
	s_{m} \;=\; \frac{1}{2} (s^{\vee}_{m} + s^{\wedge}_{m}) \spc ,
	\label{51a}
\end{equation}
which has the advantage of being Hermitian (with respect to the 
scalar product (\ref{3z})). Furthermore, we introduce the series of
operator
products
\[ b_m^< \;=\; \sum_{k=0}^\infty (-s_m \:{\cal{B}})^k
\;\;\;,\;\;\;\;\;\;
b_m \;=\; \sum_{k=0}^\infty (-{\cal{B}} \: s_m)^k \:{\cal{B}}
\;\;\;,\;\;\;\;\;\;
b_m^> \;=\; \sum_{k=0}^\infty (-{\cal{B}} \:s_m)^k \]
and set for $Q \subset N$
\[ F_m(Q,n) \;=\; \left\{ \begin{array}{ll}
	p_m & {\mbox{if $n \in Q$}} \\
	k_m & {\mbox{if $n \not \in Q$}} \end{array} \right. \spc . \]

\begin{Thm}
\label{thm1}
The relations (\ref{2tm}),(\ref{51}) uniquely determine the 
perturbation expansion for $k_m$ and $p_m$. We have the explicit 
formulas
\begin{eqnarray}
\tilde{k}_m &=& \sum_{\beta=0}^\infty (-i \pi)^{2 \beta} \; b_m^<
\:k_m\:
(b_m \: k_m)^{2\beta} \: b_m^> 
\label{52} \\
\tilde{p}_m &=& \sum_{\beta=0}^\infty \sum_{\alpha=0}^{\left[ 
\frac{\beta}{2} 
\right]} c(\alpha,\beta) \: G_m(\alpha,\beta)
\label{53}
\end{eqnarray}
with the coefficients
\begin{eqnarray}
c(0,0) &=& 1 \spc , \\
c(\alpha,\beta) &=& \sum_{n=\alpha+1}^\beta (-1)^{n+1} \:\frac{(2n-3)!!}
{n! \: 2^n} \: \left( \!\! \begin{array}{c} \beta-\alpha-1 \\ n-\alpha-1
\end{array} \!\! \right)
	\:{\mbox{for $\beta \geq 1$}}
	\label{54}
\end{eqnarray}
and the operator products
\begin{eqnarray}
 G_m(\alpha,\beta) &=& \sum_{Q \in {\cal{P}}(\beta+1) , \;\; 
\# Q=2\alpha+1}  (-i \pi)^{2\beta} \: b_m^< \:F_m(Q,1) \:b_m k_m b_m\: F_m(Q,2) \nonumber \\
&& \times \:b_m k_m b_m
	\:\cdots\: b_m k_m b_m \:F_m(Q,\beta+1)\: b_m^> \;\;\; , 
	\label{55}
\end{eqnarray}
where ${\cal{P}}(n)$ denotes the set of subsets of $\{1,\ldots,n\}$
(we use the convention $l!!=1$ for $l \leq 0$).
\end{Thm}
{\Proof}
Notice that $(i \Pdd + {\cal{B}} - m) \: b_m^< = 0$. Since all 
operator products in (\ref{52}),(\ref{55}) have a factor $b_m^<$ at the
left,
the operators $\tilde{p}_m$, $\tilde{k}_m$ are solutions of the Dirac 
equation
\[ (i \Pdd + {\cal{B}} - m) \: \tilde{p}_m \;=\; 0 \;=\;
   (i \Pdd + {\cal{B}} - m) \: \tilde{k}_m \spc . \]
Thus the theorem gives a possible perturbation expansion for $p_m$ and
$k_m$.
We must verify that the conditions (\ref{2tm}),(\ref{51}) are satisfied
and
show uniqueness.

According to (\ref{32}), the advanced and retarded Green's function 
can be written in the form
\begin{equation}
	s^{\vee}_{m} \;=\; s_{m} + i \pi \:k_{m} \;\;\;,\spc
	s^{\wedge}_{m} \;=\; s_{m} - i \pi \:k_{m} \spc .
	\label{32a}
\end{equation}
We substitute the sums (\ref{2t2}),(\ref{2t3}) into (\ref{2tm}),
\begin{equation}
	\tilde{k}_m \;=\; \frac{1}{2 \pi i} \sum_{k=0}^\infty \left(
		(- s^\vee_m \:{\cal{B}})^k \: s^\vee_m \:-\:
		(- s^\wedge_m \:{\cal{B}})^k \: s^\wedge_m \right) \spc ,
	\label{2t7}
\end{equation}
insert (\ref{32a}) and expand. This gives a sum of operator products of
the
form
\[ C_{1} \:{\cal{B}}\: C_{2} \:{\cal{B}}\: \cdots \:{\cal{B}}\: C_{l+1}
\spc {\mbox{with}} \spc C_j=k_m {\mbox{ or }} C_j=s_m \spc . \]
The contributions with an even number of factors $k_{m}$ have the same 
sign for the advanced and retarded Green's function and cancel in 
(\ref{2t7}). The contributions with an odd number of $k_{m}$'s occur 
in every Green's function exactly once and have opposite sign. Using 
the notation
\[ C_m(Q, n) \;=\; \left\{ \begin{array}{ll}
	k_m & {\mbox{if $n \in Q$}} \\
	s_m & {\mbox{if $n \not \in Q$}} \end{array} \right.
	\;\;\;, \spc Q \subset \N \spc , \]
we can thus rewrite (\ref{2t7}) in the form
\begin{eqnarray*}
\tilde{k}_m &=& \sum_{l=0}^\infty \: (-1)^l \!\!\! \sum_{ Q \in
{\cal{P}}(l+1) , \;\; \# Q \; {\mbox{\scriptsize odd}} }  (i
\pi)^{\#Q-1} \\
&& \hspace*{1cm} \times \;
	C_m(Q,1) \: {\cal{B}} \: C_m(Q,2) \: {\cal{B}} \cdots
	{\cal{B}} \: C_m(Q,l) \: {\cal{B}} \: C_m(Q,l+1) \spc.\spc
\end{eqnarray*}
After reordering the sums, this coincides with (\ref{52}).

Next we want to give the relation (\ref{51}) a mathematical sense. For
this,
we consider $m \geq 0$ as a variable mass parameter. Then we can form 
products of the operators $p_m, k_m$ by manipulating the arguments of 
the distributions in momentum space. For example, we have with 
(\ref{20})
\begin{eqnarray}
p_m(k) \: p_{m^\prime}(k) &=& (k \slsh + m) \: \delta(k^2 - m^2) \;
   (k \slsh + m^\prime) \: \delta(k^2 - (m^\prime)^2) \nonumber \\
&=& (k^2 + (m+m^\prime) k \slsh + m m^\prime) \: \delta(m^2 - 
   (m^\prime)^2) \; \delta(k^2 - m^2) \nonumber \\
&=& (k^2 + (m+m^\prime) k \slsh + m m^\prime) \: \frac{1}{2m} \:\delta(m
- 
   m^\prime) \; \delta(k^2 - m^2) \nonumber \\
&=& \delta(m-m^\prime) \: p_m(k) \spc ,
\label{56}
\end{eqnarray}
and similarly with (\ref{21}),
\begin{eqnarray}
p_m \: k_{m^\prime} &=& k_{m^\prime} \: p_m \;=\; \delta(m-m^\prime) 
\: k_m
\label{57} \\
k_m \: k_{m^\prime} &=& \delta(m-m^\prime) \: p_m \spc .
\label{58}
\end{eqnarray}
This formalism has some similarity with the bra/ket notation in 
quantum mechanics, if the position variable $\vec{x}$ is replaced by 
the mass parameter $m$.
Equation (\ref{56}) can be understood directly from the fact that 
$p_m$ are the spectral projectors of the free Dirac operator; the 
relations (\ref{57}),(\ref{58}) reflect the relative minus sign in 
$k_m$ for the states on the upper and lower mass shell. Especially one
sees
that $k_m \:k_{m^\prime} = p_m \:p_{m^\prime}$. This relation can be 
extended to the case with interaction,
\begin{equation}
	\tilde{p}_m \: \tilde{p}_{m^\prime} \;=\; \tilde{k}_m \: 
	\tilde{k}_{m^\prime} \spc ,
	\label{59}
\end{equation}
and gives a meaningful square of (\ref{51}) (we will see in a moment
that
$\tilde{k}_m \: \tilde{k}_{m^\prime}$ vanishes for
$m \neq m^\prime$). If our construction 
ensures that $\tilde{p}_m$ is a positive operator, (\ref{59}) is even 
equivalent to (\ref{51}).

We calculate the product $\tilde{k}_m \:\tilde{k}_{m^\prime}$ 
explicitly: The definitions (\ref{20}),(\ref{21}) and 
(\ref{51a}),(\ref{8b}) yield in analogy to (\ref{56}) the formulas\footnote{Note added in
January 2009: As noticed by A.\ Grotz, in~(\ref{62}) the summand $\pi^2 \delta(m-m') \,p_m$
is missing. This error is corrected in the recent paper arXiv:0901.0334 [math-ph].}
\begin{eqnarray}
p_m \: s_{m^\prime} &=& s_{m^\prime} \: p_m \;=\; {\mbox{PP}} \left(
  \frac{1}{m-m^\prime} \right) \: p_m
\label{60} \\
k_m \: s_{m^\prime} &=& s_{m^\prime} \: k_m \;=\; {\mbox{PP}} \left(
  \frac{1}{m-m^\prime} \right) \: k_m
\label{61} \\
s_m \: s_{m^\prime} &=& {\mbox{PP}} \left(
  \frac{1}{m-m^\prime} \right) \: (s_m - s_{m^\prime}) \spc ,
\label{62}
\end{eqnarray}
where ${\mbox{PP}}(x^{-1}) = \frac{1}{2} \: \lim_{0<\varepsilon 
\rightarrow 0} [ (x+i \varepsilon)^{-1} + (x-i \varepsilon)^{-1}]$ 
denotes the principal value.
As a consequence, the operator products with factors $s_m, 
s_{m^\prime}$ are telescopic, i.e.
\begin{equation}
	\sum_{p=0}^n k_m \:({\cal{B}} \:s_m)^p \: (s_{m^\prime} 
	\:{\cal{B}})^{n-p} \: k_{m^\prime} \;=\; 0 \spc {\mbox{for $n \geq 1$}}
.
	\label{63a}
\end{equation}
This allows us to easily carry out the product
$b_m^> \: b_m^<$ in the expression
\begin{equation}
	k_m \:b_m^> \:b_{m^\prime}^< \: k_{m^\prime} \;=\; 
	\delta(m-m^\prime) \: p_m \spc . 
	\label{63}
\end{equation}
With this formula, we can calculate the square of (\ref{52}) to
\begin{equation}
	\tilde{k}_m \: \tilde{k}_{m^\prime} \;=\; \delta(m-m^\prime) \:
	   \sum_{\beta_1, \beta_2 = 0}^\infty (-i \pi)^{2\beta_1 + 2\beta_2} \:
	   b_m^< \:(k_m \:b_m)^{2\beta_1} \: p_m\: (b_m \: k_m)^{2\beta_2} \:
	   b_m^> \; .
	\label{64}
\end{equation}

We could continue the proof by verifying explicitly that the product
$\tilde{p}_m \: \tilde{p}_{m^\prime}$ with $\tilde{p}_m$ according to
(\ref{53}) coincides with (\ref{64}). This is a
straightforward computation, but it is rather lengthy and not very 
instructive. We prefer to describe how the operator 
products (\ref{55}) and the coefficients (\ref{54}) can be derived.
In order to keep the proof better readable, we make some
simplifications:
Since the factors $b_m^<$, $b_m^>$ cancel similar to (\ref{63}) 
in telescopic sums, we can omit them in all formulas without changing
the multiplication rules for the operator products. Then all operator 
products have $k_m$ or $p_m$ as their first and last factor, and we 
can multiply them with the rules (\ref{56}),(\ref{57}), and (\ref{58}).
Since all these rules give a factor $\delta(m-m^\prime)$, we will in 
any case get the prefactor $\delta(m-m^\prime)$ in (\ref{64}).
Therefore we can just forget about all factors $\delta(m-m^\prime)$ 
and consider all expressions at the same value of $m$. 
Furthermore, we will omit the subscript `$_m$' and write the
intermediate
factors $b$ as a dot `.'.
After these simplifications, we end up with formal products of the form
\begin{equation}
	F_1 \:.\: F_2 \:.\: F_3 \:.\: \cdots \:.\:F_n \spc {\mbox{with}} \spc
        F_j=k {\mbox{ or }} F_j=p
	\label{41c}
\end{equation}
and have the multiplication rules
\begin{equation}
	p^2 \;=\; k^2 \;=\; 1 \;\;\;,\spc p \:k \;=\; k \:p \;=\; k \spc .
	\label{71}
\end{equation}
We must find a positive operator $\tilde{p}$ being a formal sum over 
operator products (\ref{41c}) such that
\begin{equation}
	\tilde{p}^2 \;=\; \sum_{\beta_1, \beta_2 = 0}^\infty
	(-i \pi)^{2\beta_1 + 2\beta_2} \: (k \:.)^{2 \beta_1} \:p\:
	(. \:k)^{2 \beta_2} \spc .
	\label{65}
\end{equation}
In this way, we have reduced our problem to the combinatorics of 
the operator products. As soon as we have found a solution $\tilde{p}$
of 
(\ref{65}), the expression for $\tilde{p}_m$ is
obtained by adding the subscripts `$_m$' and by inserting the factors
$b_m^<$, $b_m$, $b_m^>$.
Relation (\ref{59}) follows as an immediate consequence of (\ref{65}).

The basic step for the calculation of $\tilde{p}$ is to rewrite 
(\ref{65}) in the form
\begin{equation}
\tilde{p}^2 \;=\; p + A \spc{\mbox{with}}\spc A \;=\; 
\sum_{(\beta_1,\beta_2) \neq (0,0)} (-i \pi)^{2 \beta_1 + 2 \beta_2} \:
(k \:.)^{2\beta_1} \:p\: (. \:k)^{2\beta_2} \;\;\; .
\label{44d}
\end{equation}
The operator $p$ is idempotent and acts as the identity on $A$, 
$Ap=pA=A$. Therefore we can take the square root of $p+A$ with a
formal Taylor expansion,
\begin{equation}
	\tilde{p} \;=\; \sqrt{p+A} \;=\; p \:+\: \sum_{n=1}^\infty 
	(-1)^{n+1} \: \frac{(2n-3)!!}{n! \: 2^n} \: A^n \spc ,
	\label{74}
\end{equation}
which uniquely defines $\tilde{p}$ as a positive operator.

It remains to calculate $A^n$. If we take the $n$th power of the sum 
in (\ref{44d}) and expand, we end up with one sum over more complicated
operator products. We first consider how these operator products look
like:
The operator products in (\ref{44d}) all contain an even number of
factors
$k$ and exactly one factor $p$. The factor $p$ can be the 1st,
3rd,\ldots\
factor of the product. Each combination of this type occurs in $A$ 
exactly once. If we multiply $n$ such terms,
the resulting operator product consists of a total odd number of factors 
$p, k$. It may contain several factors $p$, which all occur at odd 
positions in the product. Furthermore, the total number of factors $p$ 
is odd, as one sees inductively. We conclude that $A^n$ consists of 
a sum of operator products of the form
\begin{equation}
	(k \:.\: k \:.)^{q_1} \;p\:.\:k\:.\; (k\:.\:k\:.)^{q_2} 
	\;p\:.\:k\:.\; (k\:.\:k\:.)^{q_3} \:\cdots\:
	(k\:.\:k\:.)^{q_{2\alpha+1}} \;p\; (.\:k\:.\:k)^{q_{2\alpha+2}}
	\label{72}
\end{equation}
with $\alpha, q_j \geq 0$. We set $\beta=2\alpha+\sum_j q_j$. Notice
that the
number of factors $p$ in (\ref{72}) is $2\alpha+1$; the total number of
factors $p, k$ is $2\beta+1$. The form of the operator product gives the 
only restriction $0 \leq 2\alpha \leq \beta$ for the choice of the
parameters 
$\alpha, \beta$.

Next we count how often every operator product (\ref{72}) occurs in the
sum:
The easiest way to realize (\ref{72}) is to form the 
product of the $\alpha+1$ factors
\begin{equation}
	\left[ (k.k.)^{q_1} \:p\: (.k.k)^{q_2+1} \right] \:
	\left[ (k.k.)^{q_3+1} \:p\: (.k.k)^{q_4+1} \right] \:\cdots\:
	\left[ (k.k.)^{q_{2\alpha+1}+1} \:p\: (.k.k)^{q_{2\alpha+2}} \right]
\;\;\; 
	\label{73}
\end{equation}
However, this is not the only possibility to factorize (\ref{72}).
More precisely, we can apply to each factor in (\ref{73}) the
identities
\begin{eqnarray*}
(k\:.\:k\:.)^q \:p\: (.\:k\:.\:k)^r &=& \left[ (k\:.\:k\:.)^q \:p
\right] \:
   \left[ p\: (.\:k\:.\:k)^r \right] \\
(k\:.\:k\:.)^q \:p\: (.\:k\:.\:k)^r &=& \left[ (k\:.\:k\:.)^s \:p
\right] \:
   \left[ (k\:.\:k\:.)^{q-s} \:p\: (.\:k\:.\:k)^r \right] \\
(k\:.\:k\:.)^q \:p\: (.\:k\:.\:k)^r &=& \left[ (k\:.\:k\:.)^q \:p\:
   (.\:k\:.\:k)^{r-s}  \right] \: \left[ p\:(.\:k\:.\:k)^s \right] \spc
.
\end{eqnarray*}
By iteratively substituting these identities into (\ref{73}), we can 
realize every factorization of (\ref{72}). Each substitution step
increases
the number of factors by one. The steps are independent in the sense
that we
can fix at the beginning at which positions in (\ref{73}) the product
shall
be split up, and can  then apply the steps in arbitrary order. There are
$(\alpha+1)+(q_1-1)+\sum_{j=2}^{2\alpha+1} q_j + (q_{2\alpha+2}-1) = 
\beta-(\alpha+1)$ positions 
in (\ref{73}) where we could split up the product (in the case $q_1=0$
or
$q_{2\alpha+2}=0$, the counting of the positions is slightly different,
but
yields the same result). Since we want to 
have $n$ factors at the end, we must choose $n-(\alpha+1)$ of these 
positions, which is only possible for $\alpha+1 \leq n \leq \beta$ and
then
gives $(\beta-\alpha-1)!/((n-\alpha-1)! \: (\beta-n)!)$ possibilities.

Combining these combinatorial factors with the constraints $0\leq 
2\alpha \leq \beta$, $\alpha+1 \leq n \leq \beta$ gives for $n \geq 1$
\begin{eqnarray}
A^n &=& \sum_{\beta=n}^\infty \sum_{\alpha=0}^{\min \left( n-1,
\left[ \frac{\beta}{2} \right] \right)}
   \left( \!\! \begin{array}{c} \beta-\alpha-1 \\ n-\alpha-1 \end{array}
\!\! 
      \right) \: \sum_{Q \in {\cal{P}}(\beta+1), \;\; \# Q = 2\alpha+1}
	  \nonumber \\
&& \hspace*{1cm} \times  \:(-i \pi)^{2\beta} \: F(Q,1) 
\:.\:k\:.\: F(Q,2) \:.\:k\:.\: \cdots \:.\:k\:.\: F(Q, \beta+1) \spc
\label{47a}
\end{eqnarray}
with $F(Q,n) = p$ for $n \in Q$ and $F(Q,n) = k$ otherwise.
Notice that the last sum in (\ref{47a}) runs over all possible 
configurations of the factors $p, k$ in the operator product (\ref{72})
for fixed $\alpha, \beta$.
We finally substitute this formula into (\ref{74}) and pull the sums 
over $\alpha, \beta$ outside. This gives the desired formula for
$\tilde{p}$.
\QED
In order to illustrate the derived formulas for $\tilde{p}$ and
$\tilde{k}$, we give the contribution up to third order in more 
detail:
\begin{eqnarray*}
\lefteqn{ \tilde{k}_m \;=\; k_m \:-\: k_m \:{\cal{B}}\:s_m \:-\: s_m 
	\:{\cal{B}}\:k_m } \\
	 && +s_m \:{\cal{B}}\: s_m \:{\cal{B}}\: k_m
	 \:+\: s_m \:{\cal{B}}\: k_m \:{\cal{B}}\: s_m
	 \:+\: k_m \:{\cal{B}}\: s_m \:{\cal{B}}\: s_m
	 \:-\: \pi^2 \:k_m \:{\cal{B}}\: k_m \:{\cal{B}}\: k_m \\
	 &&-s_m \:{\cal{B}}\:s_m \:{\cal{B}}\:s_m \:{\cal{B}}\:k_m
	 \:-\:s_m \:{\cal{B}}\:s_m \:{\cal{B}}\:k_m \:{\cal{B}}\:s_m \\
	 &&\hspace*{1.5cm}
	 -s_m \:{\cal{B}}\:k_m \:{\cal{B}}\:s_m \:{\cal{B}}\:s_m
	 \:-\:k_m \:{\cal{B}}\:s_m \:{\cal{B}}\:s_m \:{\cal{B}}\:s_m \\
	 &&+\pi^2\:s_m \:{\cal{B}}\:k_m \:{\cal{B}}\:k_m \:{\cal{B}}\:k_m
	 \:+\:\pi^2\:k_m \:{\cal{B}}\:s_m \:{\cal{B}}\:k_m \:{\cal{B}}\:k_m \\
	 &&\hspace*{1.5cm}+\pi^2\:k_m \:{\cal{B}}\:k_m \:{\cal{B}}\:s_m
	 \:{\cal{B}}\:k_m
	 \:+\:\pi^2\:k_m \:{\cal{B}}\:k_m \:{\cal{B}}\:k_m \:{\cal{B}}\:s_m
	 \:+\:{\cal{O}}({\cal{B}}^4) \\
\lefteqn{ \tilde{p}_m \;=\; p_m \:-\: p_m \:{\cal{B}}\:s_m \:-\: s_m 
	\:{\cal{B}}\:p_m } \\
	 && +s_m \:{\cal{B}}\: s_m \:{\cal{B}}\: p_m
	 \:+\: s_m \:{\cal{B}}\: p_m \:{\cal{B}}\: s_m
	 \:+\: p_m \:{\cal{B}}\: s_m \:{\cal{B}}\: s_m \\
	 &&-\frac{\pi^2}{2} \:p_m \:{\cal{B}}\: k_m \:{\cal{B}}\: k_m
	 \:-\:\frac{\pi^2}{2} \:k_m \:{\cal{B}}\: k_m \:{\cal{B}}\: p_m \\
	 &&-s_m \:{\cal{B}}\:s_m \:{\cal{B}}\:s_m \:{\cal{B}}\:p_m
	 \:-\:s_m \:{\cal{B}}\:s_m \:{\cal{B}}\:p_m \:{\cal{B}}\:s_m \\
	 &&\hspace*{1.5cm}
	 -s_m \:{\cal{B}}\:p_m \:{\cal{B}}\:s_m \:{\cal{B}}\:s_m
	 \:-\:p_m \:{\cal{B}}\:s_m \:{\cal{B}}\:s_m \:{\cal{B}}\:s_m \\
	 &&+\frac{\pi^2}{2} (s_m \:{\cal{B}}\:p_m \:{\cal{B}}\:k_m
	 \:{\cal{B}}\:k_m \:+\:p_m \:{\cal{B}}\:s_m \:{\cal{B}}\:k_m
	 \:{\cal{B}}\:k_m \\
	 &&\hspace*{1.5cm}+p_m \:{\cal{B}}\:k_m \:{\cal{B}}\:s_m \:{\cal{B}}\:k_m
	 \:+\:p_m \:{\cal{B}}\:k_m \:{\cal{B}}\:k_m \:{\cal{B}}\:s_m) \\
	 &&+\frac{\pi^2}{2} (s_m \:{\cal{B}}\:k_m \:{\cal{B}}\:k_m
	 \:{\cal{B}}\:p_m \:+\:k_m \:{\cal{B}}\:s_m \:{\cal{B}}\:k_m
	 \:{\cal{B}}\:p_m \\
	 &&\hspace*{1.5cm}+k_m \:{\cal{B}}\:k_m \:{\cal{B}}\:s_m \:{\cal{B}}\:p_m
	 \:+\:k_m \:{\cal{B}}\:k_m \:{\cal{B}}\:p_m \:{\cal{B}}\:s_m)
	 \:+\:{\cal{O}}({\cal{B}}^4) \spc .
\end{eqnarray*}
The theorem gives precise formulas for the perturbation expansion of 
the Dirac sea. Both the combinatorics of the factors $k_m, p_m, s_m$ 
and the numerical prefactors are a non-trivial result and, as far as the 
author knows, cannot be understood intuitively.

We call the perturbation expansion of this theorem the {\em{causal
perturbation expansion}}. It allows to uniquely define the Dirac sea by
\[ \tilde{P}(x,y) \;=\; \frac{1}{2} \:(\tilde{p}_m - 
\tilde{k}_m)(x,y) \spc . \]

\section{Generalization to Systems of Dirac Seas}
\label{sec4}
\setcounter{equation}{0}
In the previous section, we defined the Dirac sea for a system of 
interacting fermions of mass $m$. A realistic model, however, is 
composed of several types of fermionic particles with masses 
$m_{1},\ldots,m_{f}$. Furthermore, the fermions of zero mass may (like 
the neutrinos in the standard model) occur only as left or right handed
particles. The perturbation ${\cal{B}}$ will in general mix up the
eigenstates to different masses and will in this way describe an
interaction
of all the fermions. We will now extend the previous construction to
this
more general setting.

First we must generalize (\ref{2a}) and define a distribution 
$P(x,y)$ which describes the system in the vacuum: In order to
distinguish
the chirality of the zero-mass fermions, we introduce $(4 \times 
4)$-matrices $X_{1},\ldots,X_{f}$. For the zero-mass fermions 
$m_{j}=0$, they can be either $X_{j}=\1$, $X_{j}=\chi_{L}$, or 
$X_{j}=\chi_{R}$, where $\chi_{L\!/\!R}=\frac{1}{2}(1 \mp \gamma^{5})$ 
are the chiral projectors. For $m_{j} \neq 0$, they must coincide with 
the identity $X_{j}=\1$. The Dirac seas of the individual types of 
fermions are then described by $X_j \:\frac{1}{2}
(p_{m_j}-k_{m_j})(x,y)$. 
The remaining question is how to build up $P(x,y)$ from the individual 
Dirac seas. In view of the configuration and the interactions of the 
fermions in the standard model, one might want to use combinations of 
sums and direct sums
\begin{equation}
	P(x,y) \;=\; \bigoplus_l \sum_\alpha X_{l \alpha} \:\frac{1}{2} \: 
	(p_{m_{l \alpha}}-k_{m_{l \alpha}})(x,y)
	\label{41n}
\end{equation}
(e.g. with $l=1,\ldots,8=2(3+1)$ running over the color, lepton, and
isospin index, and with the index $\alpha=1,\ldots,3$ to distinguish the 
three fermion families. It seems reasonable to use the ordinary sum 
over $\alpha$ because the families show the same interactions).
From the mathematical point of view, however, it is easier to use only
direct sums
\begin{equation}
	P(x,y) \;=\; \bigoplus_{l=1}^{f} \: X_l \: \frac{1}{2}(p_{m_l} -
	k_{m_l})(x,y) \spc .
	\label{a}
\end{equation}
This is no loss of generality, because the ansatz (\ref{41n}) can
be obtained from (\ref{a}) by 
taking a suitable partial trace over the $l$-index (in our example, by
choosing $f=24=3 \cdot 8$ and forming the trace over the three
families).
For the perturbation expansion, we can also restrict ourselves
to the ansatz (\ref{a}), because the perturbation expansion for
(\ref{41n}) is obtained by taking the partial trace of $\tilde{P}(x,y)$
(see \cite{F3} for a more detailed discussion of this method).
Therefore we must in the following only consider a $P(x,y)$ of the form
(\ref{a}); it is called the {\em{fermionic projector of the vacuum}}.

It is convenient to use a matrix notation in the direct sum:
We set
\[ p(x,y) \;=\; \oplus_{l=1}^{f} \:p_{m_l}(x,y) \;\;\;,\spc
   k(x,y) \;=\; \oplus_{l=1}^{f} \:k_{m_l}(x,y) \]
and define the matrices
\[ X \;=\; \bigoplus_{l=1}^{f} X_l \;\;\;,\spc Y \;=\; \frac{1}{m} 
\:\bigoplus_{l=1}^{f} m_l \spc , \]
which are called {\em{chiral asymmetry matrix}} and {\em{mass matrix}},
respectively ($m$ is an arbitrary mass parameter; e.g.\ one can 
choose $m=\max_j m_j$).
Then we can write the fermionic projector as
\begin{equation}
	P(x,y) \;=\; X \: \frac{1}{2} \:(p(x,y) - k(x,y)) \spc .
	\label{81n}
\end{equation}
Since $m_j=0$ for $X \neq \1$ and because $p_{m=0}$, $k_{m=0}$ 
anti-commute with $\gamma^5$, we have alternatively
\begin{equation}
	P(x,y) \;=\; \frac{1}{2} \:(p(x,y) - k(x,y)) \: X^* \spc ,
	\label{49d}
\end{equation}
where $X^* = \gamma^0 X^\dagger \gamma^0$ is the adjoint with respect 
to the scalar product (\ref{3z}). The fermionic projector is a solution
of
the free Dirac equation
\[ (i \Pdd_{x} - m Y) \:P(x,y) \;=\; 0 \spc . \]

In order to describe the interacting system, we again insert a
differential 
operator ${\cal{B}}$ into the Dirac equation.
Thus the fermionic projector $\tilde{P}(x,y)$ is supposed to be a
solution of the Dirac 
equation
\begin{equation}
	(i \Pdd_x + {\cal{B}} - mY) \: {\tilde{P}}(x,y) \;=\; 0 \spc .
	\label{b}
\end{equation}
${\cal{B}}$ may be non-diagonal in the ``Dirac sea index'' $l$; we
assume 
it to be Hermitian with respect to the scalar product
\[ \bra \Psi \:|\: \Phi \ket \;=\; \sum_{l=1}^{f} \int 
\overline{\Psi_{l}(x)} \: \Phi_{l}(x) \; d^{4}x \spc . \]

The perturbation expansion for $k$ and $p$ can be carried out exactly as 
in the previous section: We define the advanced and retarded Green's
functions by
\[ s^\vee(x,y) \;=\; \oplus_{j=1}^f \:s^\vee_{m_j}(x,y) \;\;\;,\spc
   s^\wedge(x,y) \;=\; \oplus_{j=1}^f \:s^\wedge_{m_j}(x,y) \spc . \]
Their perturbation expansion is, in analogy to (\ref{2t2}),(\ref{2t3}),
uniquely given by
\begin{equation}
\tilde{s}^\vee \;=\; \sum_{k=0}^\infty (-s^\vee \:{\cal{B}})^k \: s^\vee
	\;\;\;,\spc
   \tilde{s}^\wedge \;=\; \sum_{k=0}^\infty (-s^\wedge \:{\cal{B}})^k \:
   s^\wedge \spc .
\label{79}
\end{equation}
\begin{Thm}
The perturbation expansion for $p$ and $k$ is uniquely determined by the 
conditions
\begin{equation}
\tilde{k} \;=\; \frac{1}{2 \pi i} \: (\tilde{s}^\vee - 
\tilde{s}^\wedge) \;\;\;,\spc
\tilde{p} \;\stackrel{\mbox{\scriptsize{formally}}}{=} 
\sqrt{\tilde{k}^2} \spc .
\label{80}
\end{equation}
We have the explicit formulas
\begin{eqnarray*}
\tilde{k} &=& \sum_{\beta=0}^\infty (-i \pi)^{2\beta} \; b^< \:k\: (b \: 
k)^{2\beta} \: b^>
\;\;\;,\spc
\tilde{p} \;=\; \sum_{\beta=0}^\infty \sum_{\alpha=0}^{\left[ 
\frac{\beta}{2} 
\right]} c(\alpha,\beta) \: G(\alpha,\beta)
\end{eqnarray*}
with
\begin{eqnarray*}
c(0,0) &=& 1 \spc , \\
c(\alpha,\beta) &=& \sum_{n=\alpha+1}^\beta (-1)^{n+1} \:\frac{(2n-3)!!}
{n! \: 2^n} \:
	\left( \!\! \begin{array}{c} \beta-\alpha-1 \\ n-\alpha-1 \end{array}
\!\!
	\right) \spc {\mbox{for $\beta \geq 1$ and}} \\
G(f,g) &=& \sum_{Q \in {\cal{P}}(\beta+1) , \;\; \# Q=2\alpha+1}
\!\!\!\!\!\!\!(-i \pi)^{2\beta} \: b^< \:F(Q,1) \:b k b\: F(Q,2) \:b
k b
	\:\cdots\: \\
\\
&&\;\;\;\;\;\;\;\;\;\;\;\;\;\;\;\;\;\;\;\;\;\;\;\;\;\; \times bkb \:  F(Q,\beta+1)\: b^> \;\; ,
\end{eqnarray*}
where ${\cal{P}}(n)$ is the set of subsets of $\{1,\ldots, n\}$
and where we used the notation
\begin{eqnarray*}
s &=& \frac{1}{2} \:(s^\vee + s^\wedge) \;\;\;,\spc
F(Q,n) \;=\; \left\{ \begin{array}{ll}
	p & {\mbox{if $n \in Q$}} \\
	k & {\mbox{if $n \not \in Q$}} \end{array} \right. \\
b^< &=& \sum_{k=0}^\infty (-s \:{\cal{B}})^k \;\;\;,\spc
b \;=\; \sum_{k=0}^\infty (-{\cal{B}} \:s)^k \:{\cal{B}} \;\;\;,\spc
b^> \;=\; \sum_{k=0}^\infty (-{\cal{B}} \:s)^k .
\end{eqnarray*}
\end{Thm}
{\Proof}
Follows exactly as Theorem \ref{thm1}.
\QED

After this straightforward generalization, we come to the more
interesting
question of how $\tilde{P}$ can be defined. Our first idea is to set
in generalization of (\ref{81n})
\begin{equation}
\tilde{P}(x,y) \;=\; X \: \frac{1}{2} \: (\tilde{p} - \tilde{k})(x,y)
\spc .
\label{84}
\end{equation}
This is not convincing, however, because we could just as well have
defined 
$\tilde{P}(x,y)$ in analogy to (\ref{49d}) by $\tilde{P} = 
\frac{1}{2} (\tilde{p}-\tilde{k}) \: X^*$, which does not coincide with
(\ref{84}) as soon as $X, X^*$ do not commute with ${\cal{B}}$.
It turns out that this arbitrariness in defining the Dirac sea reflects
a
basic problem of the causal perturbation expansion for systems with
chiral
asymmetry. In order to describe the problem in more detail, we consider
the perturbation calculation
for $k$ to first order: According to (\ref{79}),(\ref{80}), we have
\begin{eqnarray}
\tilde{k} &=& k \:-\: \frac{1}{2 \pi i} \:(s^\vee \:{\cal{B}}\: s^\vee
\:-\:
	s^\wedge \:{\cal{B}}\: s^\wedge) \:+\: {\cal{O}}({\cal{B}}^2)
	\label{81} \\
&=& k \:-\: s \:{\cal{B}}\: k \:-\: k \:{\cal{B}}\: s \:+\:
{\cal{O}}({\cal{B}}^2) \spc . \nonumber
\end{eqnarray}
This expansion is causal in the sense that $\tilde{k}(x,y)$ only 
depends on ${\cal{B}}$ in the ``diamond'' $(L^\vee_x \cap L^\wedge_y)
\cup
(L^\vee_y \cap L^\wedge_x)$, as is obvious in (\ref{81}).
It is not clear, however, how to insert the chiral asymmetry matrix into
this
formula. It seems most natural to replace all factors $k$ by $Xk$,
\begin{equation}
\tilde{(Xk)} \;=\; Xk \:-\: s \:{\cal{B}}\: Xk \:-\: Xk \:{\cal{B}}\: s
\:+\:
	{\cal{O}}({\cal{B}}^2) \spc .
\label{55m}
\end{equation}
This formula really gives a possible perturbation expansion for the 
system of Dirac seas. Unfortunately, it cannot be expressed similar to
(\ref{81})
with the advanced and retarded Green's functions, which means that the
causality
of the expansion is in general lost. In order to avoid this problem, one
might
want to insert $X$ at every factor $s, k$,
\begin{eqnarray}
\tilde{(Xk)} &=& Xk \:-\: Xs \:{\cal{B}}\: Xk \:-\: Xk \:{\cal{B}}\: Xs
	\:+\: {\cal{O}}({\cal{B}}^2) \nonumber \\
&=& Xk \:-\: \frac{1}{2 \pi i} \:(Xs^\vee \:{\cal{B}}\: Xs^\vee \:-\:
	Xs^\wedge \:{\cal{B}}\: Xs^\wedge) \:+\: {\cal{O}}({\cal{B}}^2)\spc .
	\label{83}
\end{eqnarray}
Similar to (\ref{81}), this expansion is causal. In general, however, it
does
not give a solution of the Dirac equation
$(i \Pdd + {\cal{B}} - m) \:\tilde{k}=0$,
which does not make sense.

The only way out of this dilemma is to impose that the perturbation
expansions
(\ref{55m}) and (\ref{83}) must coincide. This yields a condition on the
perturbation operator ${\cal{B}}$, which can be characterized as
follows:
We demand that
\begin{equation}
X s^\vee \:{\cal{B}}\: X s^\vee \;=\; s^\vee \:{\cal{B}}\: X s^\vee
\;=\; X s^\vee \:{\cal{B}}\: s^\vee \spc .
\label{90}
\end{equation}
Since the operator $s^\vee_{m=0}$ anti-commutes
with $\gamma^5$, we have $X s^\vee = s^\vee X^*$.
Substituting into the second equation of (\ref{90}) yields
the condition $X^* \:{\cal{B}} \;=\; {\cal{B}} \:X$.
Since $X$ is idempotent, this condition automatically implies the first
equation
of (\ref{90}).
We formulate the derived condition for the whole Dirac operator $i \Pdd
+ 
{\cal{B}}-m Y$ and thus combine it with the fact that chiral fermions
are massless (i.e. $X^* Y = Y X = Y$) and that $X$ is composed of 
chiral projectors (which implies that $X^* \Pdd = \Pdd X$).
\begin{Def}
The Dirac operator is called {\bf{causality compatible}}
with $X$ if
\begin{equation}
X^* \: (i \Pdd + {\cal{B}} - m Y) \;=\; (i \Pdd + {\cal{B}} - m Y) \: X
\spc .
\label{89}
\end{equation}
\end{Def}
In the perturbation expansion to higher order, the condition (\ref{89})
allows to commute $X$ through all operator products. Using idempotence
$X^2=X$, we can moreover add factors $X$ to the product, especially
\[ X \;C_1 \:{\cal{B}}\: C_1 \:{\cal{B}}\:  \cdots \:{\cal{B}} \:C_n
\;=\;
  XC_1 \:{\cal{B}}\: XC_1 \:{\cal{B}}\:  \cdots \:{\cal{B}} \:XC_n
\;\;\;\;\;\] {\mbox{with}}\[ \;\;\;\;\; C_j=p,\: C_j=k {\mbox{ or }} C_j=s
\;\;\; . \]
This ensures the causality of the perturbation expansion.
For a Dirac operator which is causality compatible with $X$, the
{\em{fermionic projector in the external field}} is uniquely defined by
(\ref{84}).

\section{Discussion, Outlook}
\setcounter{equation}{0}
In this paper, we gave the formal definition of the Dirac sea in the 
presence of external fields. The method differs considerably from 
earlier attempts to solve the external field problem (see e.g.\
\cite{T} and the references therein). Namely, in these previous
approaches, the 
Dirac sea was always constructed as the ``negative frequency solutions'' 
of the Dirac equation. The basic problem of this concept is that the 
notions of ``positive'' and ``negative'' frequency do not make sense in 
the case with general interaction. Therefore the construction was 
always limited to potentials which are either static
or have an only adiabatic time dependence. As shown in this paper, the 
notion of ``negative frequency states'' is not essential for the
description
of the Dirac sea. For a general definition of the Dirac sea, it must be 
given up and must be replaced by a suitable causality condition. In this 
way, it becomes possible to define the Dirac sea in the presence of 
potentials with arbitrary time dependence. Although the details of 
the perturbation expansion are a bit complicated, the basic concept 
is very simple. The construction is explicitly covariant. It puts the
usual
``hole''-interpretation of the Dirac equation on a satisfying
theoretical basis.

In order to clarify the connection to the usual definition of the 
Dirac sea, we describe how our definition simplifies in the limit of 
static potentials: If considered as multiplication operators, static 
potentials map functions of positive (negative) frequency into functions 
of positive (negative) frequency. Since $p$, $k$, and $s$ are diagonal
in
momentum space, they clearly also preserve the sign of the frequency.
Thus we have
\begin{equation}
[P^\pm,p] \;=\; [P^\pm,k] \;=\; [P^\pm,s] \;=\; [P^\pm,{\cal{B}}] 
\;=\; 0 \spc , \label{1n}
\end{equation}
where $P^\pm$ denote the projectors on the states of positive and
negative
frequency, respectively. The operators $p$ and $k$ only differ by a
relative
minus sign for the states of positive and negative frequency,
\[ P^\pm \:p \;=\; \pm \:P^\pm \:k \spc . \]
Using this relation together with (\ref{1n}), we can replace pairs of 
factors $p$ by pairs of factors $k$, e.g.
\begin{eqnarray*}
	\cdots p \:{\cal{B}} \cdots p \:{\cal{B}} \cdots &=& 
	\cdots p \:{\cal{B}} \cdots p \:{\cal{B}} \cdots \:(P^+ + P^-) \\
	 & = & P^+ (\cdots k \:{\cal{B}} \cdots k \:{\cal{B}} \cdots)
	\:+\: P^- (\cdots (-k) \:{\cal{B}} \cdots (-k) \:{\cal{B}} \cdots) \\
	 & = & \cdots k \:{\cal{B}} \cdots k \:{\cal{B}} \cdots \spc ,
\end{eqnarray*}
where the dots `$\cdots$' denote any combination of the operators $s$, 
$k$, $p$, and ${\cal{B}}$. This allows us to simplify the formula for
$\tilde{p}$ by only using exactly one factor $p$ in every operator
product.
After going through the details of the combinatorics, one obtains the
formula
\[ \tilde{p} \;=\; \sum_{b=0}^\infty (-i \pi)^{2b} \:b^< \:p\:(b 
\:k)^{2b} \:b^> \spc . \]
Thus the Dirac sea (\ref{84}) can be written as
\[ \tilde{P}(x,y) \;=\;\sum_{b=0}^\infty (-i \pi)^{2b} \:b^< 
\:\left[\frac{1}{2} \: X \:(p-k) \right]\:(b 
\:k)^{2b} \:b^> \spc . \]
This equation shows that $\tilde{P}(x,y)$ is composed of the 
negative-frequency eigenstates of the Dirac operator (notice that the 
expression in the brackets $[ \cdots ]$ is the fermionic projector of 
the vacuum and that all other factors preserve the sign of the 
frequency). Thus, for static potentials, our definition is equivalent to 
the usual concept of ``negative frequency states.''
On the other hand, this consideration illustrates in which way our
definition goes beyond the usual picture.

In order to get a better understanding of the time-dependent 
situation, we next consider a scattering process. For simplicity, we 
use the elementary framework of \cite{BD}, but our 
consideration also applies to the operator algebra and Fock space
formalism as e.g.\ described in \cite{AM}. We first recall how a 
scattering process is commonly described in the classical Dirac theory.
We assume the scattering to take place in finite time $t_0<t<t_1$.
This means that the external perturbation ${\cal{B}}$ in (\ref{1})
vanishes
outside this time interval,
\begin{equation}
{\cal{B}}(t, \vec{x}) \;=\; 0 \spc {\mbox{for}} \spc t \not \in [t_0, 
t_1] \;\;\; .
	\label{s1}
\end{equation}
We consider a solution $\tilde{\Psi}$ of the Dirac equation with 
interaction (\ref{1}). According to (\ref{s1}), $\tilde{\Psi}(t,
\vec{x})$ 
is, for $t<t_0$, a solution of the free Dirac equation. 
We uniquely extend this free solution to the whole Minkowski space and
denote 
it by $\tilde{\Psi}_{\mbox{\scriptsize{in}}}$, i.e.
\[ (i \Pdd - m) \:\tilde{\Psi}_{\mbox{\scriptsize{in}}} \;=\; 0 
\spc {\mbox{with}} \spc \tilde{\Psi}_{\mbox{\scriptsize{in}}}(t,\vec{x})
\;=\; 
\tilde{\Psi}(t, \vec{x}) \;\;\;{\mbox{for $t<t_0$}}. \]
Similarly, $\tilde{\Psi}(t, \vec{x})$ is also for $t>t_1$ a solution of
the free 
Dirac equation; we denote its extension by
$\tilde{\Psi}_{\mbox{\scriptsize{out}}}$,
\[ (i \Pdd - m) \:\tilde{\Psi}_{\mbox{\scriptsize{out}}} \;=\; 0 
\spc {\mbox{with}} \spc
\tilde{\Psi}_{\mbox{\scriptsize{out}}}(t,\vec{x}) \;=\; 
\tilde{\Psi}(t, \vec{x}) \;\;\;{\mbox{for $t>t_1$}}. \]
The wave functions $\tilde{\Psi}_{\mbox{\scriptsize{in}}}$ and
$\tilde{\Psi}_{\mbox{\scriptsize{out}}}$ are called the incoming and
outgoing
scattering states. The $S$-matrix $S$ maps the incoming scattering
states 
into the corresponding outgoing states, i.e.
\[ \tilde{\Psi}_{\mbox{\scriptsize{out}}} \;=\;
S \:\tilde{\Psi}_{\mbox{\scriptsize{in}}} \spc {\mbox{for every 
$\tilde{\Psi}$ with $(i \Pdd + {\cal{B}} - m) \:\tilde{\Psi} \;=\; 
0$}} \spc . \]
As a consequence of the Dirac current conservation, $S$ is a unitary
operator 
(with respect to the scalar product (\ref{s0})).
Using the scattering states, one can build up asymptotic Dirac seas for 
$t<t_0$ and $t>t_1$. Namely, for an observer in the past 
$t<t_0$, the bosonic potentials are zero. Thus it is natural for him 
to describe the vacuum with the free Dirac sea (\ref{2a}). If this 
Dirac sea is extended to the whole Minkowski space with external 
field, one gets the object
\[ \tilde{P}^\wedge(x,y) \;=\; \sum_{a=1,2} \int_{\sR^3} 
\tilde{\Psi}^\wedge_{\vec{k} a}(x) \:
\overline{\tilde{\Psi}^\wedge_{\vec{k} a}(y)} \: d\vec{k} \spc , \]
where the wave functions $\tilde{\Psi}^\wedge_{\vec{k} a}$ are the 
solutions of the perturbed Dirac equation whose incoming scattering 
states are the plane wave solutions $\Psi_{\vec{k} a}$,
\[ (i \Pdd + {\cal{B}} - m) \:\tilde{\Psi}^\wedge_{\vec{k} a} \;=\; 0 
\spc {\mbox{with}} \spc (\tilde{\Psi}^\wedge_{\vec{k} 
a})_{\mbox{\scriptsize{in}}} \;=\; \Psi_{\vec{k} a} \spc . \]
Accordingly, an observer in the future $t>t_0$ describes the vacuum with
the Dirac sea
\[ \tilde{P}^\vee(x,y) \;=\; \sum_{a=1,2} \int_{\sR^3} 
\tilde{\Psi}^\vee_{\vec{k} a}(x) \:
\overline{\tilde{\Psi}^\vee_{\vec{k} a}(y)} \: d\vec{k} \spc , \]
where
\[ (i \Pdd + {\cal{B}} - m) \:\tilde{\Psi}^\vee_{\vec{k} a} \;=\; 0 
\spc {\mbox{with}} \spc (\tilde{\Psi}^\vee_{\vec{k} 
a})_{\mbox{\scriptsize{out}}} \;=\; \Psi_{\vec{k} a} \spc . \]
The states $\tilde{\Psi}^\vee_{\vec{k} a}$ and 
$\tilde{\Psi}^\wedge_{\vec{k} a}$ have a more explicit form in terms 
of the perturbation series
\[ \tilde{\Psi}^\wedge_{\vec{k} a} \;=\; \sum_{n=0}^\infty 
(-s^\wedge\:{\cal{B}})^n \:\Psi_{\vec{k} a} \spc {\mbox{and}} \spc
\tilde{\Psi}^\vee_{\vec{k} a} \;=\; \sum_{n=0}^\infty 
(-s^\vee\:{\cal{B}})^n \:\Psi_{\vec{k} a} \spc , \]
as is immediately verified with (\ref{s1}) using that the support of the 
advanced and retarded Green's functions is the future and past light 
cone, respectively. The asymptotics of the Dirac seas is completely
described
by the $S$-matrix; namely
\begin{eqnarray}
	\tilde{P}^\wedge_{\mbox{\scriptsize{in}}} \;=\; P \;=\;
	\tilde{P}^\vee_{\mbox{\scriptsize{out}}} \\
	\tilde{P}^\wedge_{\mbox{\scriptsize{out}}} \;=\; S \:
	\tilde{P}^\vee_{\mbox{\scriptsize{out}}} \:S^{-1}  \\ 
	\tilde{P}^\vee_{\mbox{\scriptsize{in}}} \;=\; S^{-1}
	\:\tilde{P}^\wedge_{\mbox{\scriptsize{in}}} \:.
\end{eqnarray}\label{s3}
The physical scattering process is conveniently described with the 
two Dirac seas of the observers in the past and in the future:
If the physical system is described by $\tilde{P}^\wedge$, for 
example, the observer in the past is in the vacuum.
According to (\ref{s3}), $\tilde{P}^\wedge$ does in general not 
coincide with the Dirac sea $\tilde{P}^\vee$. This means that for the 
observer in the future, both positive frequency states are occupied 
and negative frequency states are unoccupied, so that for him the 
system contains both particles and anti-particles.
This explains the physical effect of pair creation. Other scattering
processes 
are described similarly.

The causal perturbation expansion yields a unique object $\tilde{P}$ 
describing the Dirac sea in the scattering process. $\tilde{P}$
coincides neither with $\tilde{P}^\vee$ nor with $\tilde{P}^\wedge$; 
since its construction involves both the advanced and retarded 
Green's functions, it can be considered as being an ``interpolation''
between 
$\tilde{P}^\wedge$ and $\tilde{P}^\vee$. At first sight, it might seem 
strange that the Dirac sea is now in both asymptotic regions $t<t_0$ 
and $t>t_1$ described by the same object. Namely, it was essential for 
our discussion of pair creation that the Dirac seas of the past 
and future observers were different. It might seem that by redefining 
the Dirac sea, we no longer have pair creation.
Clearly, this is not the case; all physical effects occur in the same
way 
regardless if one works with the asymptotic Dirac seas 
$\tilde{P}^\wedge$, $\tilde{P}^\vee$ or with $\tilde{P}$. This is 
because the $S$-matrix, which completely describes the 
physical scattering process, does not depend on the definition of the 
Dirac sea. Thus the choice of the definition of the Dirac sea in
the asymptotic regions is merely a matter of convenience.
This may require some explanation:
Suppose that we describe the Dirac sea with $\tilde{P}$. Then the 
asymptotic Dirac seas $\tilde{P}_{\mbox{\scriptsize{in}}}$ and
$\tilde{P}_{\mbox{\scriptsize{out}}}$ consist of both positive and 
negative frequency states. As a consequence, they are not stable; 
they tend to decay into the Dirac sea $P$ of all negative-energy 
states (this is clear physically from the fact that $P$ has lower 
energy than $\tilde{P}_{\mbox{\scriptsize{in}}}$ and
$\tilde{P}_{\mbox{\scriptsize{out}}}$). Taking this into account, one 
gets a consistent description of the physical observations.
A further complication with $\tilde{P}$ is that the current and energy 
distributions in the asymptotic regions are in general not homogeneous.
For these reasons, it is highly inconvenient to describe the 
scattering process only with $\tilde{P}$; it is much easier to work 
with $\tilde{P}^\wedge$ and $\tilde{P}^\vee$. But apart from these 
purely practical considerations, there is no reason against the 
description of the Dirac sea with $\tilde{P}$.
The great advantage of the causal perturbation expansion is that it
gives a unique definition of the Dirac sea, even in the region with
interaction
$t_0<t<t_1$. The Dirac sea is not defined with reference to an observer, 
but becomes a global object of space-time.

Our definition of the Dirac sea is the starting point for the more
technical
analysis in \cite{F4}, where all operator products are estimated and 
computed explicitly in an
expansion around the light cone. In order to further clarify the
definition of
the Dirac sea, we now qualitatively anticipate some results of
\cite{F4}.

First of all, we explain what ``causality'' of the perturbation
expansion
for the Dirac sea precisely means:
The expansion (\ref{2tm}) for $\tilde{k}_{m}(x,y)$ is causal in the 
strict sense that the perturbation operator ${\cal{B}}(z)$ only 
enters for $z$ in the ``diamond'' $z \in (L^{\vee}_{x} \cap 
L^{\wedge}_{y}) \cup (L^{\wedge}_{x} \cap L^{\vee}_{y})$.
Since $p_{m}(x,y)$ does not vanish for space-like $y-x$, its 
perturbation expansion, and consequently also the expansion of the 
Dirac sea, cannot be causal in this strict sense.
As is shown in \cite{F4}, the distribution $\tilde{P}(x,y)$ has
singularities 
on the light cone (i.e.\ for $(y-x)^{2}=0$).
It turns out that these singularities can be completely described in 
terms of ${\cal{B}}(z)$ and its partial 
derivatives along the convex line $z \in \overline{xy}$.
Our perturbation expansion is causal in this weaker sense. It is even 
uniquely characterized by this ``causality'' of the singularities on 
the light cone.

Both the operator products and the perturbation series were only treated
as formal expressions throughout this paper. We outline in 
which sense these expressions make mathematical sense:
It is shown in \cite{F4} that all operator products are well-defined
distributions if reasonable regularity conditions on ${\cal{B}}$ are
assumed.
The convergence of the perturbation expansion is a more difficult 
problem. For chiral and scalar/pseudoscalar potentials, convergence is
shown
in \cite{F4} for the formulas of the light-cone expansion by explicit
calculation. For a gravitational field,
the situation is more complicated, because the contributions
to $\tilde{P}(x,y)$ of higher order in ${\cal{B}}$ become more and 
more singular on the light cone. With a Taylor expansion of the
$\delta$-distribution
\[ \delta(x+a) \;=\; \delta(x) \:+\: a \:\delta^\prime(x) \:+\:
\frac{a^2}{2} \: \delta^{\prime \prime}(x) \:+\: \cdots \spc , \]
these contributions can be understood as describing a ``deformation'' of
the
light cone (corresponding to the diffeomorphism invariance of General 
Relativity), but the convergence has not yet been established
rigorously.

We finally remark that the fermionic projector $\tilde{P}(x,y)$ of 
section \ref{sec4} is considered in \cite{F3} as the basic physical 
object. In this context, the above construction gives a unique
characterization of $\tilde{P}$ by a perturbation ${\cal{B}}$ of the
Dirac
operator. This makes it possible to get a connection to the description
of
the interaction with classical potentials.
It turns out that this ``classical limit'' is 
completely determined by the singularities of $\tilde{P}(x,y)$ on the
light
cone. The ``causality'' of our perturbation expansion is then directly
related
to the locality and causality of the classical field equations.

\end{document}